\documentclass[showpacs,aps,superscriptaddress,prb,twocolumn,amsmath]
{revtex4}
\usepackage{amssymb}
\usepackage{graphicx}% Include figure files
\usepackage{dcolumn}% Align table columns on decimal point
\usepackage{bm}% bold math
\usepackage{hyperref}% add hypertext capabilities
\begin{document}

\preprint{APS/123-QED}

\title{First-principles study of native point defects in Bi$_2$Se$_3$}
%====================================================================
\author{L. Xue}
 \affiliation{Laboratory for Quantum Engineering and Micro-Nano Energy Technology, Xiangtan University, Xiangtan 411105, Hunan, China}
\author{P. Zhou}
 \affiliation{Laboratory for Quantum Engineering and Micro-Nano Energy Technology, Xiangtan University, Xiangtan 411105, Hunan, China}
\author{C. X. Zhang}
 \affiliation{Laboratory for Quantum Engineering and Micro-Nano Energy Technology, Xiangtan University, Xiangtan 411105, Hunan, China}
\author{C. Y. He}
 \affiliation{Laboratory for Quantum Engineering and Micro-Nano Energy Technology, Xiangtan University, Xiangtan 411105, Hunan, China}
\author{G. L. Hao}
 \affiliation{Laboratory for Quantum Engineering and Micro-Nano Energy Technology, Xiangtan University, Xiangtan 411105, Hunan, China}
\author{L. Z. Sun}
 \email{lzsun@xtu.edu.cn}
 \affiliation{Department of Physics, Xiangtan University, Xiangtan 411105, Hunan, China}
\author{J. X. Zhong}
 \email{jxzhong@xtu.edu.cn}
 \affiliation{Laboratory for Quantum Engineering and Micro-Nano Energy Technology, Xiangtan University, Xiangtan 411105, Hunan, China}

\date{\today}% It is always \today, today,

\begin{abstract}
Using first-principles method within the framework of the density functional theory, we study the influence of native point defect on the structural and electronic properties of Bi$_2$Se$_3$. Se vacancy in Bi$_2$Se$_3$ is a double donor, and Bi vacancy is a triple acceptor. Se antisite (Se$_{Bi}$) is always an active donor in the system because its donor level ($\varepsilon$(+1/0)) enters into the conduction band. Interestingly, Bi antisite(Bi$_{Se1}$) in Bi$_2$Se$_3$ is an amphoteric dopant, acting as a donor when $\mu$$_e$$<$0.119eV (the material is typical p-type) and as an acceptor when $\mu$$_e$$>$0.251eV (the material is typical n-type). The formation energies under different growth environments (such as Bi-rich or Se-rich) indicate that under Se-rich condition, Se$_{Bi}$ is the most stable native defect independent of electron chemical potential $\mu$$_e$. Under Bi-rich condition, Se vacancy is the most stable native defect except for under the growth window as $\mu$$_e$$>$0.262eV (the material is typical n-type) and $\Delta$$\mu$$_{Se}$$<$-0.459eV(Bi-rich), under such growth windows one negative charged Bi$_{Se1}$ is the most stable one.\\
\end{abstract}
 \maketitle
%++++++++++++++++++++++++++++++++++++++++++++++++++++++++++++w
\section{Introduction}
\indent The narrow-band-gap semiconductor Bi$_2$Se$_3$ (E$_g$$\sim$0.35eV)\cite{1,2} has been best known for a long time as an excellent thermoelectric material because of their unique near-gap electronic structure and high thermoelectric figure of merit ZT={S$^2$$\sigma$T}/K. \cite{3,4,5} Recently, with the development of topological insulators(TIs),\cite{6,7,8} it attracts the research attention again. TI is a new state having an energy gap in its bulk band structure and metallic helical states on its surface, which distinct from simple metal or insulator. \cite{9,10} Bi$_2$Se$_3$ is a strong three dimensional(3D) topological insulator with its surface states consisting of single Dirac cone at the $\Gamma$ point which is protected by the time-reversal symmetry from any time-reversal perturbation, such as crystal defects and nonmagnetic impurities.\cite{9,11,12} Particularly, its bulk band gap is much larger than the energy scale of room temperature, making it the most suitable candidate for the high-temperature spintronics application.\cite{9,11}\\
%++++++++++++++++++++++++++++++++++++++++++++++++++++++++++++
%++++++++++++++++++++++++++++++++++++++++++++++++++++++++++++
\indent Whether as thermoelectric material or topological insulators, the native defects in Bi$_2$Se$_3$ play important role in the material. As thermoelectric material, it needs well-defined electrical and thermal conductivities, high mobility of free current carriers, and thermoelectric power, all of them will be influenced by the presence of native defects. \cite{13,14,15} As a topological insulator, the intrinsic defects such as antisites or vacancies behave as n type dopant and consequently shift the Fermi level above the Dirac point, which makes it difficult to characterize the topological transport properties and to realize topological devices because both of them severely rely on the behavior of surface Dirac fermions.\cite{12,16,17} Hence, it is vital to investigate the nature of native defects in Bi$_2$Se$_3$. In experiments, lots of efforts\cite{13,18} have been pursued on such issue. Urazhdin et al.\cite{13} have found that Bi$_2$Se$_3$ single-crystal behaves as n type material under excess Bi or Se growth condition. They predicted that Bi$_{Se}$ antisites
contribute shallow acceptors when the samples are doped with excess Bi. However, the solubility of Bi in Bi$_2$Se$_3$ is very low and the n type doping behavior of the samples is derived from the compensating defects. While when the samples doped with excess Se, Se$_{Bi}$ antisites introduce shallow donor in the materials.\cite{13} Bludsk\'{a} et al.\cite{18} concluded that the presence of overstoichiometric Bi atoms can induce the following defects: (i) Se vacancy carrying two positive charges (V$^{+2}_{Se}$), which is attributed to the main reason for n-type Bi$_2$Se$_3$; (ii) Bi$_{Se}$ antisite carrying one negative charge (Bi$^{-1}_{Se}$); (iii) Bi vacancy supposed to carry three negative charges (V$^{-3}_{Bi}$); (iv) other complex defects. In theoretical calculations, Shuang-Xi Wang et al.\cite{19} have studied the formation energies of native point defects in Bi$_2$Se$_3$, but the study only gave the results under two extreme growing conditions: Se-rich and Bi-rich. The systematic study of the evolution of the formation energies according to growth conditions has not been reported as yet.\\
%++++++++++++++++++++++++++++++++++++++++++++++++++++++++++++
\indent To this end, in our present work, the stability and formation energies of native defects in Bi$_2$Se$_3$ are systematically investigated using first-principles method based on density function theory. The results indicate that Se1 vacancy is a double donor, Bi vacancy is a triple acceptor, and Se$_{Bi}$ is an always active n-type dopant in Bi$_2$Se$_3$. Interestingly, we find that Bi$_{Se1}$ is an amphoteric dopant. By analyzing the formation energies of all defects under different growth conditions (such as Bi-rich or Se-rich), we find that under Se-rich growth condition, Se$_{Bi}$ is the most stable one which is independent of electron chemical potential $\mu$$_e$. However, under Bi-rich growth condition, Se vacancy is the most stable defect, except for a small growth window which is defined by $\mu$$_e$$>$0.262eV (the material is typical n-type) and $\Delta$$\mu$$_{Se}$$<$-0.459eV(Bi-rich). In that growth window, one negative charged Bi$_{Se1}$ is the most stable native defect.\\
%++++++++++++++++++++++++++++++++++++++++++++++++++++++++++++
%============================================================
\begin{figure}
  \includegraphics[width=3.0in]{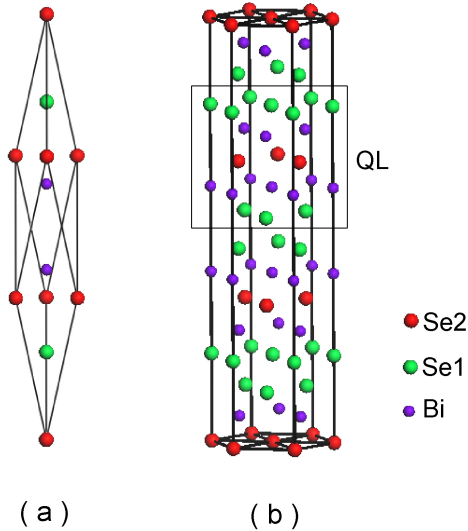}\\
  \caption{Rhombohedral primitive (a) and the hexagonal conventional unit cells (b) of Bi$_2$Se$_3$.}\label{FIG1}
\end{figure}
%============================================================
%============================================================
\begin{table}
\caption{Lattice parameters\cite{28,29,30} in our calculation for
Bi(space group R$\bar{3}$m), Se(space group P3$_1$21), and
Bi$_2$Se$_3$.}\label{tab1}
\begin{ruledtabular}
\begin{tabular}{cccccc}
System & Ref. & a({\AA}) & c/a & $\mu$$_{Bi}$ & $\nu$$_{Se}$\\
\colrule
Bi$_2$Se$_3$ & 28 & 4.138 & 6.92 & 0.399 & 0.206 \\
Bi & 29 & 4.535 & 2.611 & 0.234 &   \\
Se & 30 & 4.368 & 1.135 &   & 0.225 \\
\end{tabular}
\end{ruledtabular}
\end{table}
%============================================================
\section{Computational Method and Procedures}
\indent Bi$_2$Se$_3$ has the rhombohedral crystal structure with space group D$^5_{3d}$(R$\bar{3}$m). Its primitive unit cell contains three Se atoms and two Bi atoms, and Se atoms have two nonequivalent sites Se1 and Se2, as shown in Fig.1(a). The conventional hexagonal unit cell has a layered structure with three basic unit cells (one unit cell named a quintuple layer(QL)) weakly bound to each other by the van der Waals forces.\cite{20,21} In each QL, five atomic planes with atomic order Se1-Bi-Se2-Bi-Se1, see as fig.1(b), and the chemical bond between Bi and Se atoms is of the covalent-ionic type.\cite{22} Our calculations are performed in a 3$\times$3$\times$1 supercell with 135 atoms by using the plane-wave pseudopotential code Vienna ab initio simulation package (VASP).\cite{23,24} The core-electron interaction is modeled by the projector augmented wave method \cite{25,26} and the generalized gradient approximation of Perdew, Burke, and Ernzerhof, exchange-correlation functional \cite{27} is adopted. The plane wave cutoff energy is chosen as 240eV, and the Brillouin zone is sampled by using 5$\times$5$\times$1 Gamma-centered Mon-khorst-Pack grids. All structure parameters of Bi$_2$Se$_3$ are chosen from experimental data as listed in Table I. The energy convergent criterion is $10^{-5}$eV per unit cell. We relax all atoms without spin-orbit coupling (SOC) because the SOC only slightly affect the relaxation results and is too much time consuming. The forces on all relaxed atoms are less than 0.01eV/{\AA}. The self-consistent calculations are performed with and without spin-orbit coupling (SOC) to make a comparation. We find that the formation energies with SOC are lower than that without SOC and the difference of transition energy levels is larger than 0.04eV. In this paper, we emphasize on the results with SOC. Our calculation gives a band gap of 0.31eV, which is in good agreement with experimental and other theoretical reports.\cite{2,11,31,32}\\
%++++++++++++++++++++++++++++++++++++++++++++++++++++++++++++
\indent We consider three types of vacancy point defect in Bi$_2$Se$_3$, namely vacancy on the Se1, Se2 and Bi sublattices, denoted as V$_{Se1}$, V$_{Se2}$, and V$_{Bi}$. We sign their detail locations in Fig.2 (a). As for antisite defects, we consider Se$_{Bi}$ (substitute one Bi atom by Se atom) and Bi$_{Se1}$ (substitute one Se1 atom by Bi atom) because according to the formation energies discussed below, the formation energy of V$_{Se2}$ is 0.4eV higher than that of V$_{Se1}$, as shown in Fig.2 (b).\\
%++++++++++++++++++++++++++++++++++++++++++++++++++++++++++++
%============================================================
\begin{figure}
  \includegraphics[width=3.5in]{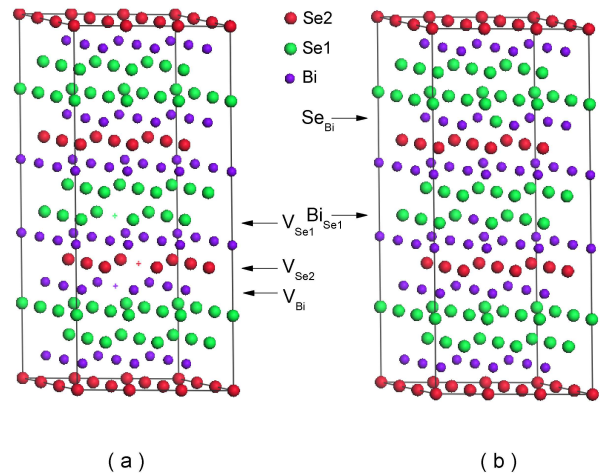}\\
  \caption{Locations of three types of vacancies (a) and two types of antisite (b) defects.}\label{FIG2}
\end{figure}
%============================================================
%=====================================================================
\begin{figure*}
\includegraphics[width=7.0in]{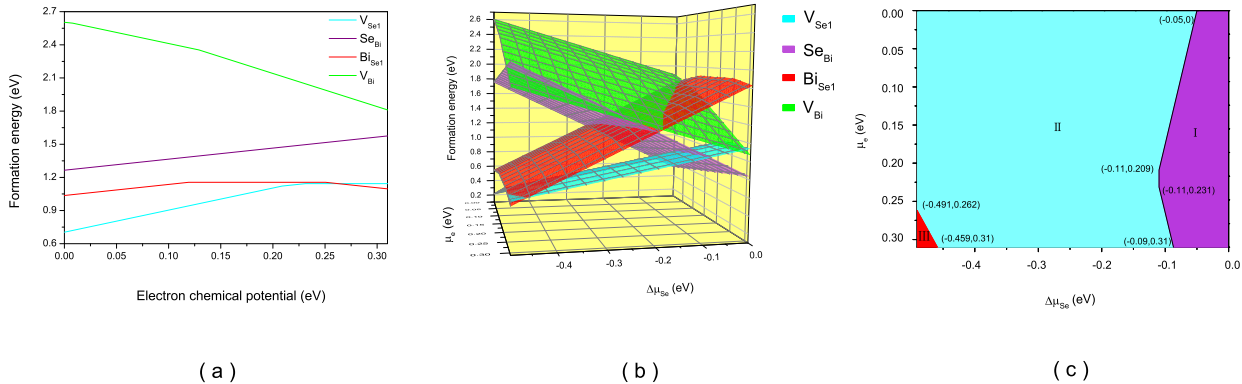}\\
\caption{(a)Formation energies of all defects as a function of the
electron chemical potential. The slope of the lines gives the charge
state. (b)Formation energy of each defect under different growth
condition, such as Se-rich($\Delta\mu_{Se}$=0eV) and
Bi-rich($\Delta\mu_{Bi}$=0eV or $\Delta\mu_{Se}$=-0.491eV). (c) is the project view figure of
figure (b). The VBM of all systems are set to zero.}\label{FIG3}
\end{figure*}
\maketitle
%=================================================================
%++++++++++++++++++++++++++++++++++++++++++++++++++++++++++++
\indent It is well known that the calculations based on GGA typically underestimate the band gap of a semiconductor and cannot give the absolute position of the defect level. In this paper, we emphasize the analysis of the defect formation energy and the transition energy levels.\cite{33,34} The formation energies of all defects in Bi$_2$Se$_3$ are given by:
%============================================================
\begin{eqnarray}\label{equ1}
H_f^{(defect, q)}&=&E_{totle}^{(defect,q)} - E_{totle}^{perfect} -
\sum_{i}\Delta n_i\mu_i \nonumber\\
&+& qE_{VBM}^{defect} + q\mu_e
\end{eqnarray}
%============================================================
where $H_f^{(defect,q)}$ is the formation energy of a defect in Bi$_2$Se$_3$ under charge state $q$. The terms, $E_{totle}^{(defect,q)}$ and $E_{totle}^{perfect}$, are the total energy of the super-cell with defect under charge state $q$ and pure Bi$_2$Se$_3$ in the same super-cell size, respectively. $\Delta$n$_i$ is the change in the number of atoms of species i, which represents Se or Bi atom in the present study. $\mu$$_i$ is the chemical potential of element $i$. $E_{VBM}^{defect}$ is the valence band maximum (VBM) of the super-cell with defect. $\mu$$_e$ is the electron chemical potential. The defect transition energy level $\varepsilon$($q$/$q^\prime$) is the electron chemical potential in Eq. (1), at which the formation energy of the defect with charge $q$ is equal to that of another charge $q^\prime$.\\
%++++++++++++++++++++++++++++++++++++++++++++++++++++++++++++
\indent To describe formation energy more accurately, we introduce two corrections as follows: i) The total energies of the charged systems should be corrected for the interaction of the charged defect with the compensating background and its periodic images. We use Makov-Payne (M-P) corrections,\cite{35} formulated as {q$^2$$\alpha$}/2{$\epsilon$L}, where L is the linear dimension of the supercell, $\epsilon$ is the static dielectric constant and $\alpha$ is Madelung constant; ii) Because the defects induce significant distortion of the band structure and fluctuation of the bandgap, the $E_{VBM}$ obtained from the first-principles method directly can not be used in Eq.(1). We adopt the correction introduced
by previous works.\cite{36,37,38,39} Firstly, we assume that the potentials in the perfect super-cell are similar to those far from a defect in a defective super-cell. Then, the average potential of the plane farthest from the defect in the defective system ($V_{a\nu}^{defect}$) and the average potential of the corresponding plane in the perfect system ($V_{a\nu}^{perfect}$) are determined. The difference of the average potentials between the perfect and defective super-cells is used to determine $E_{VBM}$ of the defective super-cell as follows:
%============================================================
\begin{eqnarray}\label{equ2}
E_{VBM}^{defect}= E_{VBM}^{perfect} + V_{a\nu}^{defect} -
V_{a\nu}^{perfect}
\end{eqnarray}
%============================================================
The first term on the right-hand side of Eq. (2) is the VBM of perfect super-cell and can be obtained by $E_{VBM}^{perfect}$ = $E_T(perfect : 0)$ - $E_T(perfect : +1)$, where E$_T$(perfect : q) is the total energy of a perfect super-cell under charge state $q$.\\
%+++++++++++++++++++++++++++++++++++++++++++++++++++++++++++++++++++
\section{Results and discussions}
\subsection{The transition energy levels}
\indent We investigate the transition energy levels of all defects quantitatively by calculating their formation energies in function of the electron chemical potential $\mu$$_e$, and the chemical potentials $\mu_i$ are considered to be the total energies of ideal elemental bulk Bi and Se crystal simply(such choice does not influence the results of the transition energy level of native defect), as shown in Fig.3(a). For Se vacancy, we consider both V$_{Se1}$ and V$_{Se2}$, and find that the formation energy of V$_{Se2}$ is 0.4eV higher than that of V$_{Se1}$. The lower formation energy of V$_{Se1}$ can be explained as a result of weak interaction between QLs. Our results indicate that the stable charge states of V$_{Se1}$ in Bi$_2$Se$_3$ are V$_{Se1}^{+2}$, V$_{Se1}^{+1}$ and V$_{Se1}^0$ as the Fermi level moves through the bandgap, which shows that V$_{Se1}$ acts as a
double donor in the systems. The results is in good agreement with the previous report.\cite{18} The donor levels $\varepsilon$(+2/+1)=0.209eV, $\varepsilon$(+1/0)=0.231eV, are 0.101eV and 0.079eV below the conduction band minimum(CBM) respectively. As for Se$_{Bi}$ in Bi$_2$Se$_3$, it is one positive charged as $\mu$$_e$ moves through the bandgap energy window. Consequently, Se$_{Bi}$ in Bi$_2$Se$_3$ behaves as a donor, and the donor level $\varepsilon$(+1/0) enter into the conduction band, indicating that it is an active n-type doping in the system. The anti-site Se protrudes outwards about 0.14{\AA} to the Se1 layer and induce the distance between the Se atoms in the Se1 layer decrease about 10.56\%. The stable charge states of isolated Bi vacancy (V$_{Bi}$) in Bi$_2$Se$_3$ as the Fermi level moves through the bandgap are V$_{Bi}^-$, V$_{Bi}^{-2}$ and V$_{Bi}^{-3}$, as shown in Fig.3(a), indicating its triple acceptors characteristics. The acceptor levels are $\varepsilon$(0/-1)=-0.07eV (the level is active in the system because it is below the valence band maximum (VBM) ), $\varepsilon$(-1/-2)=0.008eV (shallow acceptor level), and $\varepsilon$(-2/-3)=0.129eV (deep acceptor level), respectively. The results agree with the previous report.\cite{18} The last native defect considered in our present work is Bi$_{Se1}$. The results of formation energy of  Bi$_{Se1}$ as shown in Fig.3(a) indicates that the stable charge states of isolated Bi$_{Se1}$ in Bi$_2$Se$_3$ as the Fermi level moves through the bandgap are Bi$_{Se1}^+$, Bi$_{Se1}^0$ and Bi$_{Se1}^-$. Its transition energy levels are $\varepsilon$(+1/0)=0.119eV and $\varepsilon$(0/-1)=0.251eV. This means that the Bi$_{Se1}$ in Bi$_2$Se$_3$ is an amphoteric dopant, acting as a deep donor when the electron chemical potential is close to VBM (the material is typical p-type) and as a deep acceptor when the electron chemical potential is close to CBM (the material is typical n-type). Bi$_{Se1}$ will behave as a compensation dopant to the electrical type of the material. The result agrees well with the band structure of the Bi$_2$Se$_3$(111) surface with defect Bi$_{Se1}$ previously reported\cite{19} with no apparent donor or acceptor feature. Moreover, in experiment, the n-type Bi$_{Se1}$ was not mentioned and we will discuss the reasons below. As for the geometric configuration around Bi$_{Se1}$,  we find that the anti-site Bi atom protrudes outward about 0.419{\AA}, 0.425{\AA} and 0.439{\AA} relative to the average height of Se1 layer under neutral, one negative and one positive charge state, respectively. Such results are in good agreement with the experimental observation\cite{13} that there are protrusion point on the surface when Bi$_2$Se$_3$ doped with excess Bi. The protrusion can be attributed to the anti-site Bi  on the surface denoted as Se$_1$ layer in the present work.\\
%+++++++++++++++++++++++++++++++++++++++++++++++++++++++++++++++++++
%+++++++++++++++++++++++++++++++++++++++++++++++++++++++++++++++++++
\subsection{Stability of the native defect}
\indent We analyze the stability of all the native defects by calculating their formation energies under different growth environments (such as Bi-rich or Se-rich) using the method proposed by Hashibon et al.\cite{40} Using $\Delta\mu_{Bi}$($\Delta\mu_{Se}$), the difference of the chemical potential  of Bi(Se) between in the bulk Bi$_2$Se$_3$ and in ideal elemental bulk Bi(Se), the method can accurately simulate the relationship between growth condition and the formation energy of native defect. The detail process of this method is as follows:\\
\indent Firstly, the chemical potential of Bi and Se in bulk Bi$_2$Se$_3$ satisfy the relationship:
%============================================================
\begin{eqnarray}\label{equ3}
\mu_{Bi_2Se_3}^{bulk}= 2\mu_{Bi} + 3\mu_{Se}
\end{eqnarray}
%============================================================x
where $\mu_{Bi_2Se_3}^{bulk}$ is the chemical potential of one formula unit of Bi$_2$Se$_3$. The chemical potentials of Bi and Se in ideal elemental bulk are denoted as $\mu$$_{Bi}^0$ and $\mu$$_{Se}^0$, respectively. Then the the difference of the chemical potential  of Bi(Se) is defined as: $\Delta$$\mu$$_{Bi}$=$\mu$$_{Bi}$-$\mu$$_{Bi}^0$ ($\Delta$$\mu$$_{Se}$=$\mu$$_{Se}$-$\mu$$_{Se}^0$).  The Gibbs free energy in the process of formation of one formula unit of Bi$_2$Se$_3$ is given as:
%+++++++++++++++++++++++++++++++++++++++++++++++++++++++++++++++++++
%=====================================================================
\begin{figure*}
\includegraphics[width=4.5in]{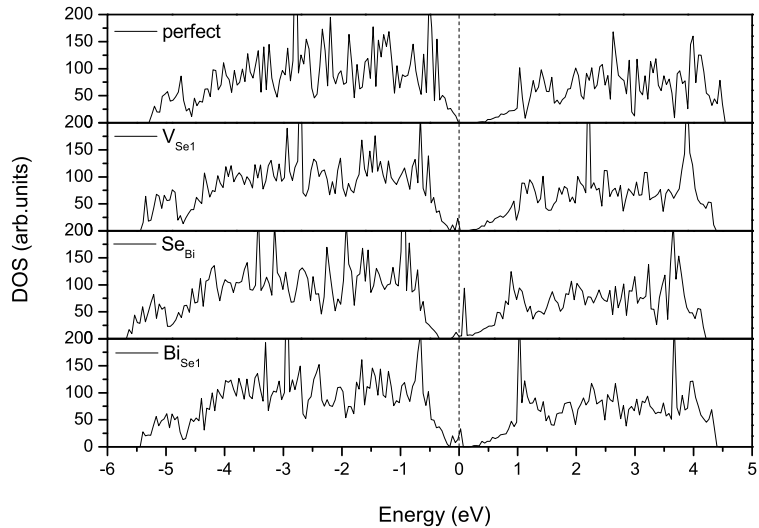}\\
\caption{Total DOS of perfect Bi$_{Se}$ system and Bi$_{Se}$
system with native point V$_{Se1}$, Se$_{Bi}$ and Bi$_{Se1}$
respectively. The Fermi level is set to zero and denoted by the
vertical dashed line.}\label{FIG4}
\end{figure*}
\maketitle
%=================================================================
%=====================================================================
\begin{table}
\caption{Formation energies (eV) of each defect at typical
Se-rich ($\Delta\mu_{Se}$=0eV) and typical
Bi-rich($\Delta\mu_{Se}$=-0.491eV or $\Delta\mu_{Bi}$=0) growth conditions
while $\mu_e$=0eV and $\mu_e$=0.31eV.}\label{tab2}
\begin{ruledtabular}
\begin{tabular}{ccccc}
 & \multicolumn{2}{c}{$\Delta\mu_{Se}$=0eV} & \multicolumn{2}{c}{$\Delta\mu_{Bi}$=0eV} \\
System & $\mu_e$=0eV & $\mu_e$=0.31eV & $\mu_e$=0eV & $\mu_e$=0.31eV \\
 V$_{Se1}$ & 0.704 & 1.144 & 0.213 & 0.653 \\
 V$_{Bi}$ & 1.867 & 1.074 & 2.604 & 1.811 \\
 Se$_{Bi}$ & 0.529 & 0.839 & 1.756 & 2.066 \\
 Bi$_{Se1}$ & 1.773 & 1.833 & 0.545 & 0.605 \\
\end{tabular}
\end{ruledtabular}
\end{table}
%============================================================
\begin{eqnarray}\label{equ4}
\Delta G(Bi_2Se_3)&=& \mu_{Bi_2Se_3}^{bulk} -2\mu_{Bi}^0-3\mu_{Se}^0 \nonumber\\
&=& 2\Delta\mu_{Bi}-3\Delta\mu_{Se}
\end{eqnarray}
%============================================================
\indent From Eq.~\ref{equ1},~\ref{equ3} and~\ref{equ4}, the formation energy formula of point defects in Bi$_2$Se$_3$ as a function of $\Delta\mu_{Se}$ can be described as:\\
%============================================================
\begin{eqnarray}\label{equ5}
H_f^{(defect, q)}=H_f^0-\frac{1}{2}\Delta G \Delta n_{Bi}-(\Delta
n_{Se}-\frac{3}{2}\Delta n_{Bi})\Delta\mu_{Se}
\end{eqnarray}
\begin{eqnarray}\label{equ6}
H_f^0 &=& E_{totle}^{(defect,q)} - E_{totle}^{perfect} - \Delta
n_{Bi}\mu_{Bi}^0 \nonumber\\&-& \Delta n_{Se}\mu_{Se}^0 +
qE_{VBM}^{defect} + q\mu_e
\end{eqnarray}
%============================================================
Similarly, the formation energy formula as a function of $\Delta\mu_{Bi}$ also can be obtained as above. Eq.~\ref{equ5} and~\ref{equ6} indicate that the formation energy is a function of $\mu_e$ and $\Delta\mu_{Se}$. In addition, Eq.(4) tells us that to form Bi$_2$Se$_3$, $\Delta\mu_{Bi}$ and $\Delta\mu_{Se}$ should meet the following condition:
%=========================================================
\begin{eqnarray}\label{equ7}
\frac{1}{2}\Delta G \leq \Delta\mu_{Bi} \leq 0
\end{eqnarray}
\begin{eqnarray}\label{equ8}
\frac{1}{3}\Delta G \leq \Delta\mu_{Se} \leq 0
\end{eqnarray}
%=========================================================
%+++++++++++++++++++++++++++++++++++++++++++++++++++++++++++++++++++
\indent The formation energies under different grown environments are shown in Fig.~\ref{FIG3}(b) and the typical results for $\mu_e$=0eV, $\mu_e$=0.31eV under $\Delta\mu_{Se}$=0eV (Se-rich) and $\Delta\mu_{Bi}$=0eV (Bi-rich) conditions are listed in Tab.~\ref{tab2}. The results as shown in Fig.~\ref{FIG3}(b) and (c) indicate that under Se-rich growth condition, the growth window denote as I in  Fig.~\ref{FIG3}(c), Se$_{Bi}$ is the most favorable native defect in the system within the full energy window of the electron chemical potential $\mu_e$, it behaves as a single donor. This means that under Se-rich growth condition Se$_{Bi}$ is the predominant native defect in the system and responsible for the n-type characteristics of the system under such growth condition. For example, when $\Delta\mu_{Se}$=0eV and $\mu_e$=0 eV, the formation energy of Se$_{Bi}$ is 0.175eV, 1.338eV and 1.244eV lower than that of V$_{Se1}$, V$_{Bi}$, and Bi$_{Se1}$, respectively. When $\mu_{Se}$=0eV and $\mu_e$=0.31eV, the formation energy of Se$_{Bi}$ is 0.305eV, 0.235eV, and 0.994eV lower than that of V$_{Se1}$, V$_{Bi}$, and Bi$_{Se1}$, respectively. When the growth window is in the region II as shown in Fig.~\ref{FIG3}(c), the most favorable native defect is V$_{Se1}$, whereas in the growth window III which is confined by $\mu_e$$>$0.262eV (the material is typical n-type) and $\Delta\mu_{Se}$$<$-0.459eV(Bi-rich) one negative charged Bi$_{Se1}$ is the most favorable one. This means that under Bi-rich growth condition, V$_{Se1}$ is the predominant native defect and it is the main reason for n-type electrical characteristics of the systems. For example, when $\Delta\mu_{Bi}$=0eV ($\Delta\mu_{Se}$=-0.491eV), the formation energy of V$_{Se1}$ according to $\mu_e$=0 eV is 0.213eV, which is 2.391eV, 1.543eV and 0.333eV lower than that of V$_{Bi}$, Se$_{Bi}$ and Bi$_{Se1}$, respectively. Only when $\Delta\mu_{Se}$$<$-0.459eV (the material is typical n-type), p-type Bi$_{Se1}$ is the dominant native defect and it behaves as a compensation dopant for the the material. The formation energy of Bi$_{Se1}$ according to $\mu_e$=0.31eV is 0.605eV, which is 0.048eV, 1.206eV and 1.461eV lower than that of V$_{Se1}$, V$_{Bi}$ and Se$_{Bi}$, respectively.\\
%+++++++++++++++++++++++++++++++++++++++++++++++++++++++++++++++++++++
\indent In experiments, the electronic properties of bulk Bi$_2$Se$_3$ crystals are usually dominated by electron donors, resulting in n-type conductivity. Based on above analysis, such n-type conductivity mainly derives from two reasons: First, the formation energy of p-type native defect V$_{Bi}$ is higher than that of n-type native defect, such as Se$_{Bi}$ and V$_{Se1}$. Second, one negative charged p-type native defect Bi$_{Se1}$ is only stable when the material is strong n-type and Bi-rich growth condition ($\Delta\mu_{Se}$$<$-0.459eV) behaving as a compensation. Moreover, we also find that one positive charged Bi$_{Se1}$ only forms when material is p-type and its formation energy is higher than that of V$_{Se1}$ which is difficult to be found in experiment.\\
%++++++++++++++++++++++++++++++++++++++++++++++++++++++++
\section{Electronic structures}
\indent The density of states (DOS) of perfect Bi$_2$Se$_3$ system and three systems with native defect of V$_{Se1}$, Se$_{Bi}$ or Bi$_{Se1}$ are shown in Fig.~\ref{FIG4}. In comparison with that of perfect system, it is clearly seen that each defect induces two defect states into the band gap of the material. For V$_{Se1}$, the defect states just below the Fermi level mainly come from p states of three nearest neighbor (NN) Bi atoms of V$_{Se1}$ and are completely filled by two electrons. The electrons filled the defect states are easily released leading the defect a double donor. The defect states of Se$_{Bi}$ mainly come from p states of four Se atoms(the Se atom substituted the Bi and its three NN Se atoms), which are half filled. The electron filled the defect state is easily released leading the defect a single donor. The defect states of Bi$_{Se1}$ mainly come from p states of four Bi atoms (the Bi atom substituted Se1 and its three NN Bi atoms), which are half filled. Depending on the conductive type of the material, either the electron filled the defect states is released leading the defect a donor or the unfilled state capture one electron leading the defect an acceptor. Bi$_{Se1}$ behaves as an amphoteric dopant in the system.\\
%+++++++++++++++++++++++++++++++++++++++++++++++++++++++++++++++++++
\indent Comparing the DOS and the formation energy results, we find that the defect levels in DOS are lower than the results of formation energy. The phenomenon may come from two possible reasons as follows: (i) the different relaxed geometries between neutral and charged system. Taking V$_{Se}$ for example, the distance between Bi atoms (nearest to the Se1 layer with vacancy) changes obviously from 3.834{\AA} for q=0 to 4.370{\AA} for q=+2; (ii)the correction of VBM for each charged system tend to make both acceptor and donor level higher. Moreover, we also note that although both Se$_{Bi}$ and Bi$_{Se1}$ induce half filled defect states around the Fermi level,  Se$_{Bi}$ acts as a single donor whereas Bi$_{Se1}$ behaves as amphoteric dopant. The reason is that the un-filled defect level above the Fermi level of Se$_{Bi}$ is higher than that of the Bi$_{Se1}$, which make it difficult to capture electron acting as p-type dopant.\\
%++++++++++++++++++++++++++++++++++++++++++++++++++++++++
\section{Conclusion}
\indent In present work, we find that V$_{Se}$ in Bi$_2$Se$_3$ is a double donor; V$_{Bi}$ is a triple acceptor; Se$_{Bi}$ is an active donor, whereas Bi$_{Se1}$ is an amphoteric dopant acting as a donor when $\mu$$_e$$<$0.119eV(the material is typical p-type) and as an acceptor when $\mu$$_e$$>$0.251eV(the material is typical n-type). The analysis of the stabilities of all defects show that under Se-rich growth condition, Se$_{Bi}$ is most stable according to all electron chemical potential $\mu$$_e$. Under Bi-rich growth condition, V$_{Se}$ is the main defect. One negative charged Bi$_{Se1}$ is most stable in a small growth window which defined by $\mu$$_e$$>$0.262eV (the material is typical n-type) and $\Delta$$\mu$$_{Se}$$<$-0.459eV(Bi-rich).\\
%++++++++++++++++++++++++++++++++++++++++++++++++++++++++
\begin{acknowledgments}
This work is supported by the National Natural Science Foundation of China (Grant Nos. 10874143, 51172191, and 10774127), the Program for New Century Excellent Talents in University (Grant No. NCET-10-0169), the Scientific Research Fund of Hunan Provincial Education Department (Grant No. 10K065) and the Hunan Provincial Innovation Foundation for Postgraduate (Grant No. CX2010B250).
\end{acknowledgments}


\end{document}